\documentclass[iop]{emulateapj}
\usepackage{graphics}
\usepackage{amsmath}

\begin{document}

\title{A mechanism for hysteresis in black hole binary state transitions}

\author{Mitchell C. Begelman and Philip J. Armitage}
\affil{JILA, University of Colorado and NIST, 440 UCB, Boulder, CO 80309-0440, USA}
\affil{Department of Astrophysical and Planetary Sciences, University of Colorado, Boulder}
\email{mitch@jila.colorado.edu}
 
\begin{abstract}
We suggest that the hysteretic cycle of black hole state transitions arises from two established properties of accretion disks: the increase in turbulent stress in disks threaded by a net magnetic field and the ability of thick 
(but not thin) disks to advect such a field radially. During quiescence, magnetic field loops are generated by the magnetorotational instability at the interface between the inner hot flow and outer thin disk. Vertical flux is advected into and accumulates stochastically within the inner flow, where it stimulates the turbulence so that $\alpha \sim 1$. The transition to a geometrically thin inner disk occurs when $L \sim \alpha^2 L_{\rm Edd} \sim L_{\rm Edd}$, and the first ``thin" disk to form is itself moderately thick, strongly magnetized, and able to advect field inward. These properties favor episodic jet production. As the accretion rate declines magnetic flux escapes, $\alpha$ decreases to $\alpha \sim 0.01 - 0.1$, and a hot inner flow is not re-established until $L \ll \ L_{\rm Edd}$. We discuss possible observational consequences of our scenario.
\end{abstract} 

\keywords{accretion, accretion disks --- binaries: close --- black hole physics --- magnetic fields --- X-rays: binaries} 

\section{Introduction} 
Accretion onto compact objects is almost always highly variable, but in black hole 
binaries the variability can be particularly dramatic. Disks in these systems exhibit 
transitions between a luminous state in which the emission contains a substantial 
thermal component, and a quiescent state dominated by non-thermal emission \citep{remillard06}. 
The luminous state, which can be accompanied by a non-thermal ``coronal" component and disk wind \citep{miller08,ponti12,king13}, is identified 
with a geometrically thin disk extending close to the innermost stable orbit \citep{shakura73}, 
while the quiescent state is consistent with the presence of a geometrically thick hot  
accretion flow \citep{rees82,narayan96} and, often, a persistent jet \citep{fender01}. 

Hot accretion requires luminosities $L \lesssim \alpha^2 L_{\rm Edd}$ \citep[where 
$L_{\rm Edd}$ is the Eddington luminosity and $\alpha$ the usual viscosity parameter:][]{esin97}, so the 
existence of a transition from a thin to a thick disk as the accretion rate wanes is not 
surprising. More puzzling is the behavior at the start of a new outburst. Rather than 
backtracking in a diagram of spectral hardness versus luminosity, the system remains in a 
non-thermal state up to a much higher luminosity --- in some cases, approaching the Eddington limit. Transient radio emission from jets is seen as the disk enters the luminous thin disk state \citep{fender04}. 
The origin of this hysteretic cycle, and much else besides, is not understood. Candidate explanations include 
ideas as diverse as the conjectured dependence 
of $\alpha$ on magnetic Prandtl number \citep{balbus08}, and instabilities associated 
with disk warps \citep{nixon14}.

In this Letter, we describe a mechanism for hysteresis based on established 
properties of accretion flows in which angular momentum is transported by the magnetorotational 
instability (MRI). In common with previous authors \citep{meier05,reynolds06,igumenshchev09,dexter13}, we argue that dynamically significant net fields thread the inner disk while it is hot and geometrically thick. 
Net fields increase the strength of the MRI \citep{hawley95}, leading to $\alpha \sim 1$ 
for net fields whose ratio of gas to magnetic pressure $\beta \lesssim 10^2$ \citep{bai13}. 
With such efficient transport the density is low, and this 
precludes transition to a thin disk until $L \sim L_{\rm Edd}$. 
Once a thin disk has formed and the accretion rate has dropped significantly below Eddington, 
the net flux is able to diffuse out radially through the disk \citep{lubow94}, and $\alpha$ 
drops. The hot inner flow is then re-established at a much lower $L \ll L_{\rm Edd}$. To 
complete the cycle, net flux is regenerated in the inner thick disk at the 
interface with the adjacent thin disk. The MRI acting near the interface generates loops of 
vertical field, some fraction of which accumulate stochastically within the hot flow.

\section{Hysteresis from cyclic flux accumulation}

\begin{figure*}
\includegraphics{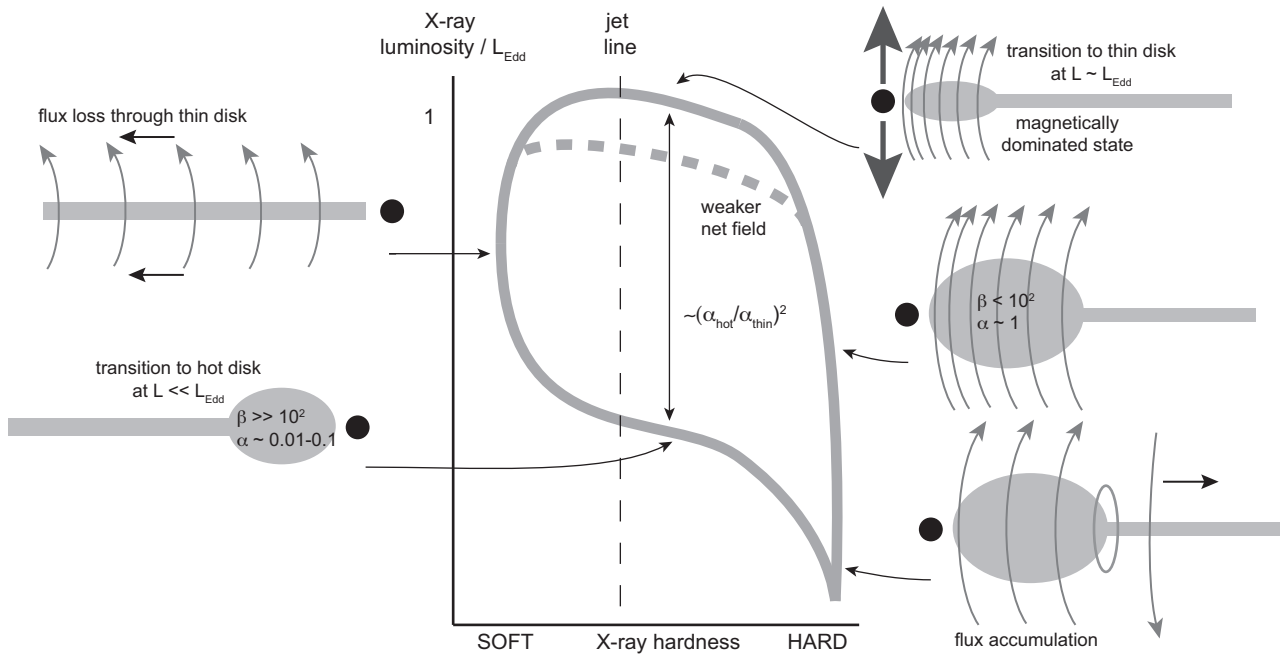}
\caption{Illustration of the observed evolution of black hole binary disks in $L_X$ versus spectral hardness 
\citep[simplified from][]{fender04}, and how that relates to the assumed evolution of net magnetic fields in our 
model. The luminosity of the upper branch of the transition depends on the net accumulated flux, which can vary from cycle to cycle. }
\label{fig1}
\end{figure*}

Early numerical simulations of the magnetorotational instability \citep{balbus98} 
demonstrated that although the MRI can act as a dynamo -- sustaining turbulence in the 
absence of mean fields -- the strength of MRI turbulent transport increases when a net 
field is present. \cite{hawley95}, analyzing a set of unstratified local simulations that 
included net vertical or toroidal field, found that the saturated magnetic pressure $\langle B^2/8\pi \rangle$ 
and viscosity parameter $\alpha$ both scaled linearly with the net field $B_0$. Recent 
work confirms this trend. Without net field, the MRI generates an $\alpha \simeq 0.01$ \citep{davis10}, 
but this is stimulated by any net vertical field (characterized by a mid-plane ratio of gas to 
magnetic pressure $\beta$) with 
\begin{equation}
 \beta = \frac{\rho c_s^2}{B_z^2 / (8 \pi)} \lesssim 10^5.
\end{equation} 
\cite{bai13} find that $\alpha \simeq 0.08$ for $\beta = 10^4$, and that 
$\alpha \gtrsim 1$ for $\beta = 10^2$. Disk winds are also 
produced as the net field increases \citep{suzuki09,lesur13,bai13,simon13}, though their strength 
and structure remain somewhat uncertain.

Hot accretion flows cannot exist for $L \gtrsim \alpha^2 L_{\rm Edd}$, because radiative 
cooling is too efficient \citep{esin97}. The hysteresis observed in black hole disk state 
transitions is thus consistent with the hot state possessing a substantial net flux, 
leading to a high $\alpha \sim 1$ and a transition luminosity approaching the 
Eddington limit, while the thin disk state is weakly magnetized with a lower $\alpha$ and 
a small transition luminosity. (The hot flow could be ``magnetically dominated" with 
$\beta \sim 1$, as in the model of \citet{igumenshchev09}, but need not be provided 
that $\beta$ is low enough that $\alpha \sim 1$.) The simulations suggest that the gap 
in transition luminosities could be as large as a factor of $10^3$ to $10^4$, though less extreme 
values are possible if flux accumulation is inefficient or the thin disk retains some net flux. Thin disks 
are unable to advect vertical field inward \citep{lubow94}, so it is plausible to expect 
that an initially magnetized thin disk of scale $r_{\rm out}$ will expel its net field on the diffusion 
time $t_\eta \sim r_{\rm out}^2 / \eta$ associated with the turbulent diffusivity $\eta$.

If the thin disk expels net flux, how is it regenerated? We argue that flux accumulation 
in the vicinity of the black hole occurs due to the action of an MRI disk dynamo \citep{king04} at 
the interface between an inner hot flow and an outer thin disk . Vertical loops of magnetic field, with radial scales ranging up to $\sim h$ and zero net flux, are created at random in both the thick and thin disk regions. Loops formed deep within the inner hot flow are advected inward and do not contribute to flux accumulation, whereas loops entirely within the thin disk diffuse away.  But close to the interface, field loops can be created where one footpoint is trapped in the inner hot zone while the other escapes into the thin disk.  As the loop opens up and the footpoints lose causal contact, the inner hot flow is left with an element of net magnetic flux, of random sign, that is uncompensated.  Inward advection of these uncorrelated elements of magnetic flux can lead to the stochastic buildup of a large net magnetic flux close to the black hole.  

Because the scale height of the thin disk is so small ($h/r \sim 0.01$), flux accumulation due to the MRI in the thin disk is probably too slow to be of interest.  We therefore focus on the hot flow near the interface at radius $r$, assuming $h/r \sim 1$, a characteristic random vertical field strength corresponding to $\beta_0$, and a time scale for random changes in flux $t_{\rm dyn}$.  The mid-plane gas pressure $P \sim \alpha^{-1}(\Omega / r) \dot{M}$, where 
$\dot{M}$ is the accretion rate and $\Omega$ the angular velocity. During each dynamo cycle $t_{\rm dyn} = k \Omega^{-1}$, $N$ patches appear carrying positive or negative flux of magnitude  
\begin{equation}
 |\Delta \Phi|_N \sim {\pi r^2 \over 2 N}  \left( \frac{8\pi P}{\beta_0}  \right)^{1/2}  
\end{equation} 
each.  If a fraction $f$ of these patches escapes to the thin disk and diffuses away, a net increment of flux $\sim  (f N)^{1/2} |\Delta \Phi|_N$ is acquired by the hot flow, where it can be advected inward \citep{igumenshchev08,beckwith09,cao11}.  If these flux parcels accumulate as a random walk, the hot flow will have enough net field to attain $\beta \sim 1$ after a time
\begin{equation}
 t \sim {4 N \over f} \beta_0 k  \left( \frac{r}{r_g} \right)^{3/2} \frac{GM}{c^3}.
\end{equation} 
Here $r_g = GM/c^2$. MRI turbulence exhibits moderately well-defined dynamo cycles, with $k \sim 10^2$ 
\citep[e.g.][]{oneill11}, and it is reasonable to assume flux cells of scale up to $\sim r$ in a thick disk. It is not 
clear, however, how weakly magnetized the thick disk is when it first forms, or how readily flux is separated 
at the interface with the thin disk. For $M = 10 \ M_\odot$ and $\beta_0 = 10^3$ (a weak enough field that 
$\alpha$ would be less than unity) we have $t \sim 6 \times 10^5 (N/f)(r / 10^3 r_g)^{3/2} \ {\rm s}$. Provided that 
$N/f$  is of order $\sim 100$ or less this suggests that a dynamically important field can be regenerated during 
the quiescent phase of X-ray binaries, in far less time than the interval between outbursts. We also note that 
the magnetic field of the donor star will thread (at least) the outer disk, and that even if this ambient flux 
cannot be advected it may act as a boundary condition affecting the build up of flux within the hot flow.

Figure~\ref{fig1} illustrates how these elements may contribute to hysteresis in state transitions. 
In common with other authors, we assume that the 
action in the inner disk is sourced by large-scale variations in the accretion rate originating 
from further out. These variations are consistent \citep{coriat12} with being due to 
thermal instability associated with the ionization of 
hydrogen \citep{hoshi79,meyer81} . For typical parameters, thermal instability originates 
at radii $r \sim 10^5 r_g$ and can be considered independent of the physics of state transitions. 
The quiescent state, with a hot inner accretion flow, corresponds to the thermally unstable 
outer disk also being quiesent, with a low accretion rate. The extent to which magnetic flux 
can accumulate in the inner flow will depend on the interval between outbursts, and on the 
details of the flux accumulation process. Moderately strong constraints can be deduced from 
the observation that steady jets are present during quiescence, if one assumes that these are powered by the 
black hole spin \citep{blandford77}. In this case a simple estimate for the 
power is $P_{BZ} = (\kappa/4 \pi c) \Phi^2 \Omega_H^2$, with $\Phi$ the flux threading 
the horizon, $\Omega_H$ the horizon angular velocity, and $\kappa \simeq 0.05$ a constant 
\citep[for better estimates, informed by numerical simulations, see e.g.][]{tchekhovskoy10}. 
The efficiency of the Blandford-Znajek process in a thick disk is then $\epsilon \propto \alpha^{-1} \beta^{-1}$, 
with a prefactor typically substantially below unity even for a rapidly spinning hole. Hence, 
even though we only require $\beta \lesssim 10^2$ to reach $\alpha \sim 1$, stronger fields 
would be needed to produce jets. A magnetically dominated accretion flow is also 
required in models where Quasi-Periodic Oscillations (QPOs) have frequencies linked  
to the magnetospheric radius \citep{igumenshchev09}. These arguments are not 
watertight (jets, for example, could also be produced from disk winds), but they 
motivate consideration of models where the hot state resembles a magnetically dominated 
accretion flow \citep{dexter13}.

For $\alpha \sim 1$ the transition to a thermal state occurs at $L \sim L_{\rm Edd}$. How this 
transition occurs is unclear, but if it takes place on the rapid thermal time scale the immediate 
consequence will be to reduce the disk pressure, so that the already strong magnetic fields 
become even more dominant. The high accretion rate, moreover, ensures that the nominally 
``thin'' disk will in fact have a substantial $(h/r)$ \citep[see e.g. the slim disk solutions 
of][]{abramowicz88}. These properties suggest that the newly formed inner disk will be able to 
advect flux inward even more efficiently than the precursor hot flow, with the angular 
momentum transport probably being dominated by wind loss rather than by the MRI 
\citep{lubow94b,bai13}. At the same time strong toroidal field will escape the disk 
vertically on a short time scale. The transient jets observed to form during the thick to 
thin transition could be a consequence of this flux expulsion \citep{shibata86}, with an 
independent physical origin to the steady jets seen during quiescence.

Even after a thin disk has formed and a quasi-steady accretion state has been attained, 
the disk close to the black hole will still be strongly magnetized. 
It is likely that a corona will be present, and inflow through the 
magnetically dominated atmosphere of the disk will in turn affect (and slow) flux 
expulsion \citep{lovelace09,guilet12}. Over time flux will escape but if initially 
$\beta \sim 1$ there will be a period during which $\beta$ remains $\le 10^2$, 
$\alpha$ is of the order of unity, and the disk is close to the boundary between 
thin and hot accretion flows. Fluctuations in the accretion rate, which could 
originate at large scales in the disk \citep[e.g. due to irradiation coupling between 
the inner disk and the thermally unstable zone:][]{king98} or from instabilities 
in flows where radiation pressure is significant, will therefore be able to trigger 
the observed reversions of the inner disk to a hot state. Only after $\beta \ge 10^2$ 
will $\alpha$ start to drop significantly. The inner disk will then remain stably 
in a thin disk state until a large decline of the accretion rate has occurred. 
Finally the accretion rate and luminosity drop to $L \simeq \alpha^2 L_{\rm Edd}$, 
with $\alpha = 0.02 - 0.1$, closer to the value expected for a zero net-field disk.  This is the level needed 
to re-establish a hot and initially weakly magnetized inner flow, completing the cycle.

\section{Discussion}
We have argued that the hysteresis observed in the cycle of black hole X-ray binary state 
transitions has an elementary explanation; the hot accretion flow present in the low / hard 
state is more strongly magnetized than the geometrically thin disk that forms subsequently. 
The strength of MRI disk turbulence increases with field strength and this, in turn, results 
in the thick-to-thin transition occurring at a higher accretion rate than the thin-to-thick 
transition. The MRI can be boosted even if the disk is threaded by vertical fields that 
reverse sign with radius \citep{sorathia10,beckwith11}, but we have emphasized the 
simplest possibility that the hot flow has an organized net field. A net field can 
build up during quiescence, and leak away during outburst, because only thick 
disks (or perhaps thin disks with strong coronae) are able to advect net fields 
radially. The details of this flux cycle are the most uncertain aspect of our 
model, and numerical simulations of the evolution of magnetic fields near the 
interface between thick and thin disks would be valuable.

Observations indicate that the luminosity at the upper transition fluctuates considerably from source to source, and from cycle to cycle in the same source. These fluctuations might reflect the stochastic nature of flux accumulation, which could lead to a distribution of net magnetic field strengths and corresponding fluctuations in the value of $\alpha$. The luminosity at the lower transition, typically  $\sim 10^{-2} L_{\rm Edd}$, shows much less scatter \citep{maccarone03,kalemci13}, suggesting that on this branch the net field is too weak to affect $\alpha$. 

How and where thermal state transitions originate in black-hole binary accretion flows is uncertain, hence we are unable to propose a quantitative timeline for the cycle.  However, a clear prediction of our model is that the ratio of luminosities between the cold-to-hot and hot-to-cold transitions is sensitive to the ratio of $\alpha$-parameters between the highly magnetized and weakly magnetized states.  Timescales that depend on $\alpha$ should reflect this dependence. 

\citet{igumenshchev09} and \citet{dexter13} have already proposed that the hot state of black hole X-ray binaries 
is strongly magnetized \citep[in their case a ``magnetically arrested disk'':][]{narayan03}, 
and the phenomenology of jets and variability in these systems appears to be consistent 
with this hypothesis \citep{dexter13}. \citet{igumenshchev09} further suggested 
that QPO frequencies might be linked to the magnetospheric radius of such a flow.
In our model, the dynamical importance of the net field varies cyclically 
as the inner disk transitions from a hot flow to a thin disk and back again. This suggests 
observational implications beyond those described in \cite{dexter13}. Although coronal activity is possible whenever the MRI operates in a disk, its signatures may depend on the value of $\beta$.  The disk winds that are observed in absorption during the thin disk state \citep{miller08,ponti12,king13} may owe their prominence to being launched mainly by coronal activity (i.e., the buoyancy of turbulent fields generated by the MRI), in contrast to faster, lower-density winds launched magneto-centrifugally by the net fields expected to thread the disk prior to and immediately following the thick-to-thin transition \citep{suzuki09,bai13}.

An intriguing possibility is that the collapse of the highly magnetized hot flow to a thin disk may trap an extremely strong, even suprathermal toroidal field generated by the MRI \citep{begelman07}.  The presence of such a field would thicken the disk, decreasing the viscous inflow time \citep{johansen08}, and may also lead to a large color correction in the thermal component of the spectrum, for which observational evidence may already exist \citep{salvesen13,reynolds13}.

\acknowledgments

We thank Jeff McClintock and Greg Salvesen for valuable discussions and comments on the manuscript. We acknowledge support from NASA's Astrophysics Theory Program under grant NNX11AE12G.

\end{document}